\documentclass[pdflatex,sn-mathphys-num, iicol]{sn-jnl}% Math and Physical Sciences Numbered Reference Style
\usepackage{graphicx}%
\usepackage{multirow}%
\usepackage{amsmath,amssymb,amsfonts}%
\usepackage{amsthm}%
\usepackage{mathrsfs}%
\usepackage[title]{appendix}%
\usepackage{xcolor}%
\usepackage{textcomp}%
\usepackage{manyfoot}%
\usepackage{booktabs}%
\usepackage{algorithm}%
\usepackage{algorithmicx}%
\usepackage{algpseudocode}%
\usepackage{listings}%

\usepackage[frozencache,cachedir=.]{minted}

\theoremstyle{thmstyleone}%
\theoremstyle{thmstyletwo}%

\theoremstyle{thmstylethree}%

\begin{document}

\title[Article Title]{LLMs and Fuzzing in Tandem: A New Approach to
Automatically Generating Weakest Preconditions}


\author*[1]{\fnm{Daragh} \sur{King}}\email{kingd6@tcd.ie}

\author[1]{\fnm{Vasileios} \sur{Koutavas}}\email{vasileios.koutavas@tcd.ie}

\author[2]{\fnm{Laura} \sur{Kovacs}}\email{{laura.kovacs@tuwien.ac.at}}

\affil*[1]{\orgdiv{Department of Computer Science and Statistics}, \orgname{Trinity College Dublin}, \orgaddress{\city{Dublin}, \country{Ireland}}}

\affil[2]{\orgdiv{Faculty of Informatics}, \orgname{TU Wien}, \orgaddress{\city{Vienna}, \country{Austria}}}


\abstract{
The weakest precondition (WP) of a program describes the largest set of initial states from which all terminating executions of the program satisfy a given postcondition. The generation of WPs is an important task with practical applications in areas ranging from verification to run-time error checking. This paper proposes the combination of Large Language Models (LLMs) and fuzz testing for generating WPs. In pursuit of this goal, we introduce \emph{Fuzzing Guidance} (FG); FG acts as a means of directing LLMs towards correct WPs using program execution feedback. FG utilises fuzz testing for approximately checking the validity and weakness of candidate WPs, this information is then fed back to the LLM as a means of context refinement. We demonstrate the effectiveness of our approach on a comprehensive benchmark set of deterministic array programs in Java. Our experiments indicate that LLMs are capable of producing viable candidate WPs, and that this ability can be practically enhanced through FG.}




\maketitle

\section{Introduction}\label{Intro}

Pre- and postconditions are fundamental formalisms for specifying program behaviour, originating from the seminal work of Floyd~\cite{Floyd67} and Hoare~\cite{hoare-logic}. 
Beyond Hoare-logic itself, these formalisms are a cornerstone in many software engineering, verification, and debugging techniques; these include design-by-contract~\cite{Meyer92}, abstract interpretation~\cite{CousotCousot77}, program refinement~\cite{refinement}, symbolic execution~\cite{king70}, and runtime assertion checking~\cite{Rosenblum95}.

The postcondition of a program method specifies the desired property of the program state upon termination. On the other hand, the precondition specifies the required property of the initial state for the method to behave as intended.\footnote{In total correctness specifications, a precondition also guarantees termination of the method; in this paper we restrict our attention to programs that terminate for any input.}
Knowing the pre- and postconditions of a method suffice to ensure its correctness to external calling methods -- thereby enabling compositional program reasoning.

There can be many pre- and postconditions for a given method $m$, but finding the \emph{right} ones is crucial for effective compositional reasoning.
If $m$'s precondition is too permissive (allowing too many initial states) or its postcondition too restrictive (rejecting too many final states), reasoning about $m$'s behaviour may fail. Conversely, if $m$'s precondition is too restrictive or postcondition too permissive, reasoning about methods calling $m$ may fail.

This need for precision has led to the notion of a \emph{weakest precondition} (WP), i.e.~the most permissive precondition for which $m$'s executions lead to final states satisfying its postcondition.\footnote{And its dual, the strongest postcondition.} 
The weakest precondition calculus~\cite{wp-calculus} was the first manual method for deriving the weakest precondition from a method's code and desired postcondition.
Subsequent techniques have sought to automatically derive preconditions, either sufficient (though often not weakest)
\cite{abstract-debugging93,footprint-analysis07,CousotEtal02,procedure-summaries07,modular-assertion-checking08,astorga2019learning,cousot2011precondition,gehr2015learning,Moy08,padhi2016data,sankaranarayanan2008dynamic,SeghirEtal13,sumanth2024maximal,ZhouEtal21,sumanth2024weakest} or necessary ones~\cite{CousotEtal13}. However, since the problem is undecidable in the presence of loops, user intervention remains an essential fallback.

% Generating WPs presents a significant technical challenge; this is especially true when the programs being considered contain common programming constructs like loops and recursion. These features  impose the burden of reasoning about inductive properties such as loop invariants~\cite{kovacs2013first, furia2014loop}

% A terminating program's weakest precondition (WP) defines the input space from which all executions achieve the desired postcondition. In other words, when the initial state of a terminating program adheres to the WP, it is guaranteed that the program will behave correctly -- according to its given postcondition, at least. The practicality of such a precondition is evident and wide-ranging; WPs can find application to fields such as runtime error-checking~\cite{weimer2004finding}, verification~\cite{leucker2009brief} and compositional symbolic execution~\cite{fragoso2019javert}.

% Generating WPs presents a significant technical challenge; this is especially true when the programs being considered contain common programming constructs like loops and recursion. These features  impose the burden of reasoning about inductive properties such as loop invariants~\cite{kovacs2013first, furia2014loop}. Though this task is difficult (and undecidable, in general), there exists a rich variety of methods tackling this issue~\cite{sumanth2024weakest, sumanth2024maximal, ernst2007daikon, padhi2016data}. 

Studies and tools regarding the generation of WPs (and their requisite loop invariants) typically reside on the more mathematical side of computer science, relying on formal methods. On the other end of the spectrum lie more heuristic-based, data driven fields, such as fuzz testing~\cite{manes2019art} and machine learning~\cite{jordan2015machine}. Fuzz testing generates pseudo random inputs for a program under test in an attempt to trigger buggy or unwanted behaviour. Fuzz testing has seen far-reaching success~\cite{zhu2022fuzzing}, and is actively deployed in industry~\cite{bounimova2013billions}. While fuzz testing has grown in popularity, machine learning has given rise to LLMs~\cite{naveed2023comprehensive}. 

In recent years, LLMs have found extensive and successful applications to software engineering, including code generation~\cite{chen2022codet}, code repair~\cite{jiang2023impact} and invariant inference~\cite{pirzada2024llm}. Although LLMs suffer from inherent drawbacks (e.g. stochastic outputs, hallucinations, etc.), it is natural to question the extent to which these techniques can contribute to the inherently manual process of WP generation. 
% In this regard, the application of LLMs could complement existing formal-method tools and potentially reduce the need for user involvement.
% However, zero-shot querying LLMs for WP inference is unreliable. Their effectiveness depends on factors such as their (often unknown) training data, inherent probabilistic behaviour, and susceptibility to producing so-called hallucinations---incorrect responses presented as truth.

This paper broaches a \emph{new frontier in the automatic generation of WPs}. First, we employ LLMs to act as generators for candidate WPs. Second, we introduce Fuzzing Guidance (FG) as a means for refining these candidate WPs and steering LLMs towards generating correct WPs. FG relies upon two distinct phases of fuzz testing to validate and ensure the weakness of an LLM-generated WP; the results of these fuzzing phases are then fed back to the LLM as a means of context refinement. The benefits of this combined approach are:
\begin{itemize}
    \item \emph{Superior performance in WP generation}: The combination of LLMs and FG can greatly improve the quality/correctness of generated WPs when compared to zero-shot prompting; this is especially true for (typically) cheaper non-reasoning models (e.g. GPT-4o~\cite{gpt-4o-ref}), enabling them to achieve comparable performance to their (typically) more expensive reasoning counterparts (e.g. O4-mini~\cite{o4-mini-ref}).
    \item \emph{Increased scope}: Unlike previous works in WP inference that rely on formal methods, our use of LLMs for generating WPs has no inherent limitation in the programs it can handle. This increases the potential scope and use of WPs in practice. Indeed, our experiments include programs which fall outside the scope of previous work -- these include variants of problems revolving around the use of binary search and sorting (see the supplementary material link for example-four in Section~\ref{supp-mat}).
    \item \emph{Model-driven evolution}: Although the general performance progress of LLMs is not guaranteed, whatever evolutions do occur in the future will be readily reflected in our approach. Relying on LLMs enables this background-improvement to materialise in our method for WP generation.
\end{itemize}

Section~\ref{Methods} describes the methodology of FG; outlining a high-level structure of the programs used, the prompts utilised, and the constituent phases of fuzz testing. Following this, Section~\ref{Motivating Example} depicts a motivating example; here, GPT-4o and FG interact to produce the correct WP for a particular program.

Section~\ref{Evaluatio-wp} details our experimental analysis. We introduce the four benchmark sets of Java programs used in our work, and outline the LLM-FG configurations utilised for generating  WPs. Detailing our results, we explore if LLMs can generate sensible WP candidates and whether FG can aid them in doing so by enabling iterative phases of context refinement.

Proceeding to the tail-end of the paper, we provide a discussion of our work in Section~\ref{Discussion} -- detailing the relevant threats to validity and the limitations of our approach. Section~\ref{Related Work} overviews related work. Section~\ref{Future Work} outlines future directions, and Section~\ref{Conclusions} concludes.

\begin{figure}[t!]
\centering
%\includegraphics[width=9cm, height=3cm]{images/Untitled Diagram-2.pdf}
%\fbox{%
\includegraphics[width=\columnwidth,trim=530 520 100 60]{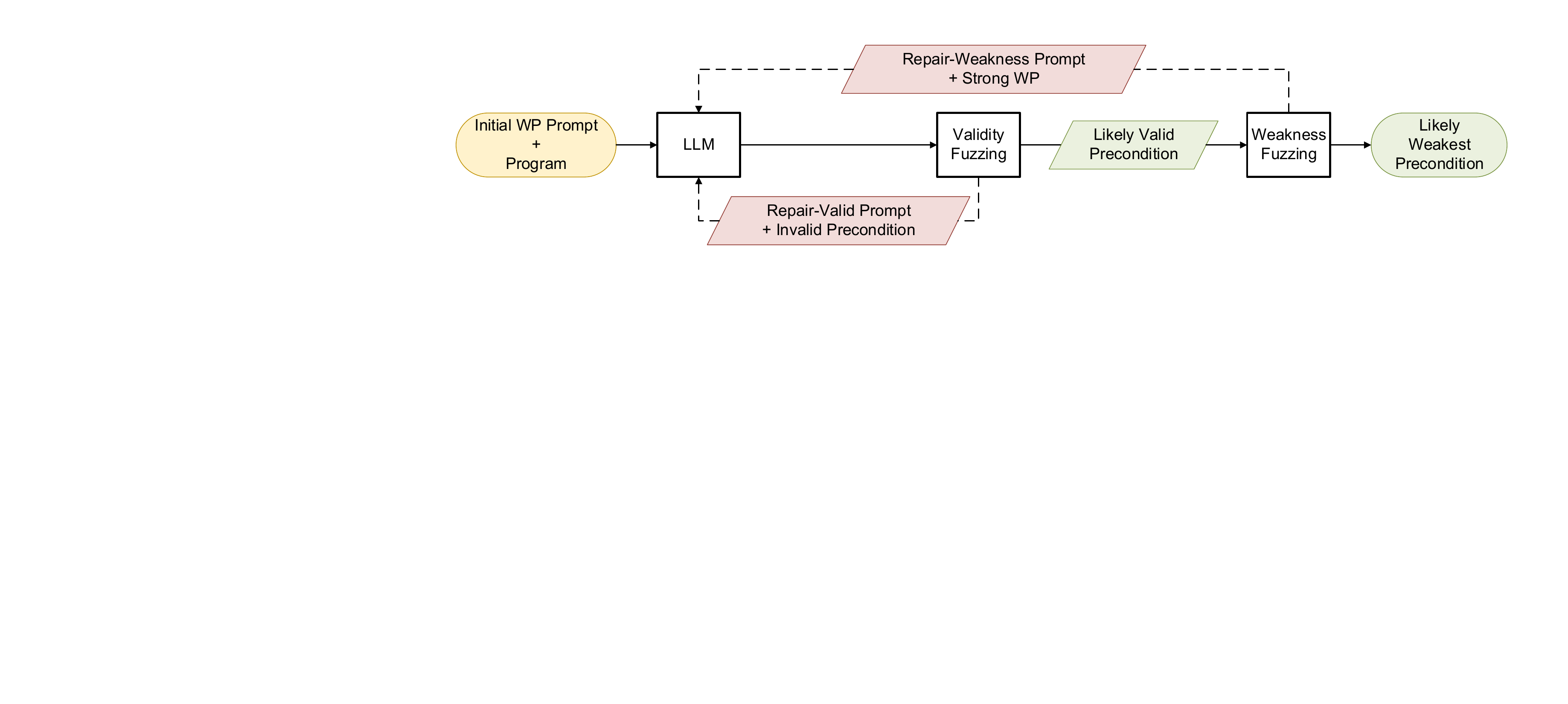}
%}
\caption{Fuzzing Guidance Workflow.}
\label{fig:fuzzing-guidance-workflow}
\end{figure}

\section{Method}\label{Methods}

This section  outlines our methodology for the generation of Weakest Preconditions (WPs) using LLMs and Fuzzing Guidance (FG). The overarching workflow for this process can be seen in Fig~\ref{fig:fuzzing-guidance-workflow} -- the fuzzing phases and prompts depicted are detailed in the Section~\ref{fuzz-guidance-phases} and Section~\ref{prompting} respectively.

\subsection{Input Programs}\label{input-programs}
The input programs considered for this work follow a common high-level structure (an example can be seen in Fig~\ref{fig:program-example} in Section~\ref{Motivating Example}). Specifically, the programs considered are implemented in Java and feature a deterministic looping computation over arrays; these programs consist of:
\begin{itemize}
    \item A single `static int' method named `foo' which accepts three arrays $a, b, c$ as inputs.
    \item A postcondition implemented in code which checks a desired property for some subset of the input arrays; an exception is thrown if the postcondition is not satisfied.
    \item A successful return value of $0$.
\end{itemize}

The looping computations contained within each program range over a varied set of tasks, e.g. array copying, arithmetic, sorting, search, etc. For example, the program in Fig~\ref{fig:program-example} of Section~\ref{Motivating Example}, copies elements of array $a$ into array $b$.

Additionally, it is worth noting that each program's single method is named `foo' to prevent any unnecessary direction being given to the LLM being prompted.

\begin{figure}[t!]
\centering
\begin{Verbatim}[frame=single, fontsize=\tiny]
Context: You have the following Java program containing a method named 'foo'. 
You must determine the weakest precondition for 'foo' so that:
    - The method’s postcondition is always satisfied.
    - No exceptions or errors are reachable.

Task A: Identify this weakest precondition.
Task B: Create a public Java method named 'precondition' that checks this 
weakest precondition. 

Adhere to the following requirements:
    - It must accept the same arguments as 'foo'.
    - It must return 'false' if the precondition is not satisfied and 'true' 
    otherwise.
    - It may include a for loop if necessary.
    - Before this method, provide a single brief comment describing the 
    precondition in pseudo-code.
    - Integrate this 'precondition' method into the original Java program and 
    return the complete updated program.

Output Format: Return only the updated Java program (do not include any other 
additional text), which includes
    - The original 'foo' method - this should be unchanged.
    - The newly added 'precondition' method.

Provide no additional explanations beyond the program code and the required 
comment. Reason through your solution internally.
\end{Verbatim}
\caption{LLM prompt to generate WPs, i.e. the initial-WP prompt.}
\label{fig:init-prompt}
\end{figure}

\subsection{Fuzzing Guidance Phases}\label{fuzz-guidance-phases}
FG utilises the JQF fuzz testing tool~\cite{padhye2019jqf} to carry out two distinct phases of fuzz testing for candidate WPs. These phases are termed \emph{validity-fuzzing} and \emph{weakness-fuzzing}; the outputs of which act as feedback to the LLM being utilised for WP generation.

\subsubsection{Validity-fuzzing}\label{val-fuzzing}

This phase of fuzzing is employed to help check that a candidate WP is valid. A candidate WP is valid if all initial program states which satisfy the WP wind up satisfying the program's given postcondition after it is executed. 

Validity-fuzzing attempts to generate pseudo-random initial states which adhere to a candidate WP, but which do \textbf{not} satisfy a program's postcondition; i.e.~they cause the program to throw an exception. If such an input is found then we can conclude that the candidate WP is not valid; otherwise we can posit that the WP is \emph{likely-valid}.

\subsubsection{Weakness-fuzzing}\label{weak-fuzzing}
Given a WP that is already deemed to be likely-valid, weakness-fuzzing attempts to ensure that the same WP is \emph{likely-weakest}. A candidate WP is weakest if it accounts for all initial program states which result in the program's postcondition being satisfied after it is executed.

Weakness-fuzzing attempts to generate pseudo-random initial states which do \emph{not} adhere to a candidate WP, but which do satisfy a programs postcondition; i.e.~no exception is thrown. If such an input is found then we can conclude that the candidate WP is not weakest; otherwise we can posit that the WP is likely-weakest.

\begin{figure}[t!]
\centering
\begin{Verbatim}[frame=single, fontsize=\tiny]
You have a Java program that contains:
    - A method called 'foo', which may reach an error state under certain 
    inputs.
    - A method called 'precondition', intended to represent the weakest 
    precondition for 'foo'.

The current implementation of 'precondition' is incorrect. Specifically, there 
is a known set of input values for which 'precondition' returns 'true' 
(indicating the precondition is satisfied), but passing these same inputs to 
'foo' actually triggers an error state. The aforementioned inputs follow:
[Insert Inputs]

Task:
    1. Analyze the existing 'foo' method and the incorrect 'precondition' 
    method.
    2. Determine why the current 'precondition' method fails to exclude the 
    problematic inputs.
    3. Rewrite or adjust 'precondition' so that it correctly represents the 
    true 
    weakest precondition for 'foo', ensuring that when 'precondition' returns 
    'true', 'foo' will not reach an error with the same inputs.

Goal:
Provide the corrected 'precondition' method so that 'foo' never encounters an 
error when the corrected 'precondition' returns 'true'.

Output Format: Return only the updated Java program (do not include any other 
additional text), which includes
    - The original 'foo' method - this should be unchanged.
    - The corrected 'precondition' method.

Provide no additional explanations beyond the program code and the required 
comment. Reason through your solution internally.
\end{Verbatim}
\caption{LLM prompt to repair or replace invalid WPs, i.e. the repair-validity prompt.}
\label{fig:repair-validity}
\end{figure}

\subsection{Prompting}\label{prompting}
Our method for generating WPs relies on three underlying prompts for LLM interactions; these will be referred to as the \emph{initial-WP prompt} (Fig.~\ref{fig:init-prompt}), the \emph{repair-validity prompt} (Fig.~\ref{fig:repair-validity}), and the \emph{repair-weakness prompt} (Fig.~\ref{fig:repair-weakness}).

The initial-WP prompt tasks the LLM with WP generation for a particular program. In isolation (i.e.~when FG is not employed) this prompting method can be viewed as a zero-shot prompting strategy~\cite{schulhoff2024prompt}. The outcomes to this zero-shot strategy are revealed through the results for the \textit{GPT-4o} and \textit{O4-mini} experimental configurations in Section~\ref{Evaluatio-wp}.

The repair-validity and repair-weakness prompts expect to deal with an already existing candidate WP. The repair-validity prompt provides the LLM with inputs which highlight that the current candidate WP is not valid (as derived through validity-fuzzing, Section~\ref{val-fuzzing}); the LLM is then tasked with repairing the current WP or replacing it with a newly generated WP. Similarly, the repair-weakness prompt provides the LLM with inputs which highlight that the current candidate WP is not the weakest possible (as derived through weakness-fuzzing, Section~\ref{weak-fuzzing}); the same task of WP repair or generation is then also put forth to the LLM.

Besides intention and functionality, there are additional aspects to our prompts which are worth discussing. Each of our prompts ends with `Reason through your solution internally.'. This is done to encourage the LLM to utilise a chain-of-thought~\cite{schulhoff2024prompt} process that does not reveal these `thoughts' in the corresponding output; this makes the parsing of LLM's responses more straightforward. One might argue that this restriction on the LLM's outputs might affect how they can be interpreted; we mitigate this issue through the use of the following phrase in our initial-WP prompt -- `provide a single brief comment describing the precondition in pseudo-code.'. This addendum ensures that 1) there is an easily-read reasoning behind the LLM's response, and 2) there exists a counter-point to the code generated when it does not align with expectations.

\begin{figure}[t!]
\centering
\begin{Verbatim}[frame=single, fontsize=\tiny]
You have a Java program that contains:
    - A method called 'foo', which may reach an error state under certain 
    inputs.
    - A method called 'precondition', intended to represent the weakest 
    precondition for 'foo'.

The current implementation of 'precondition' is incorrect. Specifically, there 
is a known set of input values for which 'foo' returns successfully and 
'precondition' returns 'false' - indicating that the inputs do not satisfy the 
given precondition, but still result in a succesful execution of 'foo'. 
This implies that the precondition defined by the 'precondition' method may 
not be weakest. The aforementioned inputs follow:
[Insert Inputs]

Task:
    1. Analyze the existing 'foo' method and the incorrect 'precondition' 
    method.
    2. Determine why the current 'precondition' method fails to correctly 
    account for the above inputs.
    3. Rewrite or adjust 'precondition' so that it correctly represents the 
    true weakest precondition for 'foo'.

Goal:
Provide the corrected 'precondition' method so that 'foo' never encounters 
an error when the corrected 'precondition' returns 'true', and that the 
corrected 'precondition' method truly represents the weakest precondition for 
'foo'.

Output Format: Return only the updated Java program (do not include any other 
additional text), which includes
    - The original 'foo' method - this should be unchanged.
    - The corrected 'precondition' method.

Provide no additional explanations beyond the program code and the required 
comment. Reason through your solution internally.
\end{Verbatim}
\caption{LLM prompt to repair or replace strong WPs, i.e. the repair-weakness prompt.}
\label{fig:repair-weakness}
\end{figure}

\subsection{Fuzzing Guidance Cycles}\label{fuzzing-cycles}
The outputs derived from the individual validity- and weakness-fuzzing phases (Section~\ref{fuzz-guidance-phases}) are combined with their respective repair prompts (Section~\ref{prompting}) in an effort to steer the LLM towards generating higher-quality candidate WPs in terms of validity and weakness respectively. This describes the essence of FG, but a little more needs to be said when considering how it is used in practice -- we must define how the two phases of FG combine to define an FG-cycle. An FG-cycle consists of:
\begin{enumerate}
    \item A maximum of $X$ iterations of validity-fuzzing and repair.
    \item A single iteration of weakness-fuzzing and repair.
\end{enumerate}
Step 2 follows the successful completion of Step 1 (cf.~Fig.~\ref{fig:fuzzing-guidance-workflow}). Multiple FG-cycles can be used in practice to generate WPs.

\begin{figure}[t!]
    \centering
    \begin{minted}
    [
    fontsize=\footnotesize
    ]
    {java}

public static int foo(int[] a, int[] b, int[] c) {
    int N = a.length;
    if (N == 0 || b.length != N) {
        throw new RuntimeException();
    }

    int i = 0;
    while (i < N) {
        b[i] = a[i];
        i++;
    }

    int[] b_clone = b.clone();
    sort(b_clone);

    i = 0;
    while (i < N) {
        if (b[i] != b_clone[i])
            throw new RuntimeException();
        i++;
    }

    return 0;
}
    \end{minted}
    \caption{Example program $P$ from the `Sorting' benchmark set (Section~\ref{benchmark-sets}).}
    \label{fig:program-example}
\end{figure}

\section{Motivating Example}\label{Motivating Example}

This section outlines the generation of a correct weakest precondition using an LLM (GPT-4o~\cite{gpt-4o-ref}) and Fuzzing Guidance (FG). We will do so for an example program $P$ (Fig.~\ref{fig:program-example}) from our `Sorting' benchmark set (Section~\ref{benchmark-sets}). (See the supplementary material link for example-one in Section~\ref{supp-mat}).

Before proceeding, it is worth re-noting that the comments in the following code snippets are generated by the LLM in question (Section~\ref{prompting}).

\subsection{Initial Weakest Precondition Generation}
First, we prompt the LLM to generate its initial candidate WP by appending program $P$ to the initial-WP prompt (Fig.~\ref{fig:init-prompt}). This results in the following `precondition' method being generated (i.e. the initial candidate WP) in Fig.~\ref{fig:initial-wp-gend}.

\begin{figure}[ht!]
    \centering
    \begin{minted}
    [
    fontsize=\footnotesize
    ]
    {java}

// Precondition: 'a' and 'b' must have the same 
// non-zero length, 
// and 'c' should be at least of length 'a'.
public static boolean precondition
(int[] a, int[] b, int[] c) {
    return a.length > 0 && a.length == b.length 
    && c.length >= a.length;
}
    \end{minted}
    \caption{Initial candidate WP generated for $P$ (Fig.~\ref{fig:program-example}).}
    \label{fig:initial-wp-gend}
\end{figure}

\subsection{Ensuring Validity}\label{ensuring-validity}
Given the initial candidate WP generated (Fig.~\ref{fig:initial-wp-gend}) for program $P$, we can now proceed to checking its validity using validity-fuzzing (Section~\ref{val-fuzzing}). Validity-fuzzing quickly highlights an issue with the given candidate; the following inputs satisfy the candidate WP (i.e. `precondition' in Fig.~\ref{fig:initial-wp-gend} returns true) but ultimately result in a failing execution of $P$.

\begin{minted}
    [
    fontsize=\footnotesize
    ]
    {java}

int[] a = {1401919545, 267222264, 1358618764};
int[] b = {739288021, 776171394, -594227544};
int[] c = {427235608, -506959485, 1997723943,
          -498207265, 733576341};
\end{minted}

The program $P$, along with the invalid candidate WP and the preceding inputs, are integrated into the repair-validity prompt (Fig.~\ref{fig:repair-validity}); using this prompt the LLM is tasked with amending the invalid candidate WP. This results in Fig.~\ref{fig:second-wp-gend}. Upon inspection, this `precondition' method appears to be a valid candidate for $P$'s WP; another round of validity fuzzing provides evidence that this is so. Now, we must attempt to see if this candidate WP is likely-weakest.

\begin{figure}[ht!]
    \centering
    \begin{minted}
    [
    fontsize=\footnotesize
    ]
    {java}

// Precondition: 'a' and 'b' must have the same 
// non-zero length, 
// 'c' should be at least of length 'a',
// and 'a' must be sorted in non-decreasing order 
// to ensure 'b' remains sorted.
public static boolean precondition
(int[] a, int[] b, int[] c) {
    if (a.length == 0 
    || a.length != b.length 
    || c.length < a.length) {
        return false;
    }
    
    for (int i = 0; i < a.length - 1; i++) {
        if (a[i] > a[i + 1]) {
            return false;
        }
    }

    return true;
}
    \end{minted}
    \caption{
    A likely-valid but too-strong candidate WP for $P$ (Fig.~\ref{fig:program-example}).}
    \label{fig:second-wp-gend}
\end{figure}

\subsection{Ensuring Weakness}
Through validity-fuzzing, we now have a candidate WP (Fig.~\ref{fig:second-wp-gend}) that we believe to be a valid precondition for the program $P$. Our goal now is to check the weakness of the candidate WP with weakness-fuzzing (Section~\ref{weak-fuzzing}). Weakness-fuzzing proceeds to highlight a problem with the current candidate WP; the following inputs do \emph{not} satisfy the WP (i.e. i.e. `precondition' in Fig.~\ref{fig:second-wp-gend} returns false) but result in a successful execution of $P$:

\begin{minted}
    [
    fontsize=\footnotesize
    ]
    {java}

int[] a = {-1286467063};
int[] b = {0};
int[] c = {};

\end{minted}

By accounting for the preceding inputs and inspecting $P$ again, we can see that the restriction on the input array $c$ in the current candidate WP (Fig.~\ref{fig:second-wp-gend}) is unnecessary. Therefore, through the processes of validity-fuzzing and weakness-fuzzing, we can determine that our current candidate WP is likely a valid precondition for $P$, but that it is demonstrably not the weakest (i.e. it is too strong) because of the unnecessary condition on the length of \texttt{c}.

The program $P$, along with the strong candidate WP and the preceding weakness-fuzzing outputs, are integrated into the repair-weakness prompt (Fig.~\ref{fig:repair-weakness}); this prompt tasks the LLM with amending the strong WP -- resulting in the candidate WP of Fig.~\ref{fig:second-last-wp-gend}.

\begin{figure}[ht!]
    \centering
    \begin{minted}
    [
    fontsize=\footnotesize
    ]
    {java}

// Precondition: 'a' and 'b' must have the same 
// non-zero length, 
// and 'c' must have a length that is 
// irrelevant to the execution success of 'foo'. 
// Additionally, 'a' naturally 
// becomes sorted when it has identical elements.
public static boolean precondition
(int[] a, int[] b, int[] c) {
    return a.length > 0 && a.length == b.length;
}
    \end{minted}
    \caption{
    A step backwards for FG? Another candidate WP for $P$ (Fig.~\ref{fig:program-example})}
    \label{fig:second-last-wp-gend}
\end{figure}

Immediately we can observe that there appears to be something amiss with the latest candidate WP (Fig.~\ref{fig:second-last-wp-gend}). It appears simpler than the candidate in Fig.~\ref{fig:second-wp-gend}, and more akin to the invalid candidate depicted in Fig.~\ref{fig:initial-wp-gend} (although it does do away with the unnecessary restraints on the array $c$ mentioned previously). We must undergo validity fuzzing again to ensure that the quality of the candidate WP has not regressed.

\subsection{Ensuring Validity and Weakness Again}
The current candidate WP (Fig.~\ref{fig:second-last-wp-gend}) must undergo validity-fuzzing again. As before in Section~\ref{ensuring-validity}, validity-fuzzing finds a set of inputs which highlight the invalidity of the current candidate (these inputs are omitted for brevity). Once again, the repair of the current candidate WP must take place with the repair-validity prompt. This process results in the following candidate WP:

\begin{figure}[ht!]
    \centering
    \begin{minted}
    [
    fontsize=\footnotesize
    ]
    {java}

// Precondition: 'a' and 'b' must have the same 
// non-zero length, 
// and 'a' must be sorted in non-decreasing order,
// since 'b' is a direct clone of 'a' before 
// sorting the clone.
public static boolean precondition
(int[] a, int[] b, int[] c) {
    if (a.length == 0 || a.length != b.length) {
        return false;
    }
    
    for (int i = 0; i < a.length - 1; i++) {
        if (a[i] > a[i + 1]) {
            return false;
        }
    }

    return true;
}
    \end{minted}
    \caption{
    A likely-valid and -weakest WP for $P$ (Fig.~\ref{fig:program-example}).}
    \label{fig:final-wp-gend}
\end{figure}

Given the latest candidate WP in Fig.~\ref{fig:final-wp-gend}, validity-fuzzing and weakness-fuzzing both succeed in succession; therefore this candidate is likely-valid and likely-weakest. Through manual inspection we can see that this is, in fact, the correct WP for program $P$.

\subsection{Conclusion}
Overall, we can see from this example how FG can enable the generation of correct WPs. Through several bouts of back-and-forth between LLM interations and different phases of fuzz testing, we arrived at a candidate WP that is correct.

\subsection{Additional Remarks}
Notice that $P$ (Fig.~\ref{fig:program-example}) calls an external method `sort', and that no additional information was provided to the LLM about this method. Despite this, the approach taken here has arrived at the correct WP for $P$ -- from this, we can conclude that LLMs can reason about unknown but well-understood methods like `sort'. 

The above phenomena is of practical benefit in terms of prompting, but there is also a danger of missing important information regarding the behaviour of these unspecified method calls. If such a concern exists, one could augment the code with the pre- and post-conditions of these methods, allowing the LLM reason more robustly in a compositional manner.

\begin{table*}[t!]
\centering
\tiny
\renewcommand{\arraystretch}{1.2}
\caption{Resulting statistics for each benchmark set and configuration, obtained across $k=5$ iterations.}
\begin{tabular}{|p{2cm}|p{2.2cm}|r|r|r|r|}
\midrule
\textbf{Configuration} & \textbf{Benchmark} & \textbf{\# Programs} & \shortstack{\textbf{\# Correct WPs} \\ \textbf{Min–Max (Avg., Avg.\%)}} & \shortstack{\textbf{FG Usage} \\ \textbf{Min–Max (Avg.)}} & \shortstack{\textbf{FG Success} \\ \textbf{Min–Max (Avg.)}} \\ \midrule\midrule

\multirow{4}{*}{\shortstack{\textit{\textbf{GPT-4o}}}} & Existential & 20 & 10 – 12 (11, 55.00 \%) & -- & -- \\ 
                                                       & Universal & 36 & 24 – 27 (26.2, 72.78 \%) & -- & -- \\ 
                                                       & Sorting & 8 & 2 – 4 (2.8, 35.00 \%) & -- & -- \\ 
                                                       & Search & 6 & 4 – 5 (4.2, 70.00 \%) & -- & -- \\ \midrule\midrule

\multirow{4}{*}{\shortstack{\textit{\textbf{GPT-4o-FG}}}} & Existential & 20 & 16 – 17 (16.6, 83.00 \%) & 5 – 7 (6) & 5 – 6 (5.6) \\ 
                                                         & Universal & 36 & 32 – 34 (33, 91.67 \%) & 6 – 10 (7.8) & 6 – 9 (6.8) \\ 
                                                         & Sorting & 8 & 4 – 6 (5.4, 67.50 \%) & 4 – 6 (4.6) & 2 – 4 (2.6) \\ 
                                                         & Search & 6 & 5 – 6 (5.8, 96.67 \%) & 1 – 2 (1.8) & 1 – 2 (1.6) \\ \midrule\midrule

\multirow{4}{*}{\shortstack{\textit{\textbf{O4-mini}}}} & Existential & 20 & 18 - 19 (18.6, 93.00 \%) & -- & -- \\ 
                                                        & Universal & 36 & 34 - 35 (34.8, 96.67 \%) & -- & -- \\ 
                                                        & Sorting & 8 & 7 - 8 (7.8, 97.50 \%) & -- & -- \\ 
                                                        & Search & 6 & 5 - 6 (5.4, 90.00 \%) & -- & -- \\\midrule\midrule

\multirow{4}{*}{\shortstack{\textit{\textbf{O4-mini-FG}}}} & Existential & 20 & 20 - 20 (20, 100 \%) & 1 – 2 (1.4) & 1 – 2 (1.4) \\ 
                                                           & Universal & 36 & 36 - 36 (36, 100 \%) & 1 – 2 (1.2) & 1 – 2 (1.2) \\ 
                                                           & Sorting & 8 & 8 - 8 (8, 100 \%) & 0 – 1 (0.2) & 0 – 1 (0.2) \\ 
                                                           & Search & 6 & 6 - 6 (6, 100 \%) & 0 – 1 (0.6) & 0 – 1 (0.6) \\ \midrule

\end{tabular}
\label{tab:overall_results}
\end{table*}

\section{Evaluation of Weakest Precondition Generation}\label{Evaluatio-wp}
The core focus of this work is to deploy LLMs and fuzz testing towards the task of generating weakest preconditions (WPs); the method employed here is outlined in detail in Section~\ref{Methods} -- in brief, LLM prompting for WP generation is augmented with Fuzzing Guidance (FG) in order to steer the LLM toward generating correct WPs. In this section we  outline our results pertaining to this goal, comparing the results of WP generation with and without FG. In assessing and presenting these results we are primarily concerned with the three following criteria: \textit{Correctness}, \textit{Stability}, and the \textit{Contribution of FG}. 

\textit{Correctness}: for a particular program, the correctness of a generated WP is defined in relation to the program's ground-truth WP. These ground-truth WPs are known a priori (through manual derivation) and the comparison with generated WPs is performed manually; if a generated WP is deemed semantically equivalent to its ground-truth counter part then it is correct.

\textit{Stability}: we complete $k$ iterations of WP generation for each benchmark set and experimental configuration. This allows us to look for consistency in the WPs generated and assess the stability of both the LLM outputs and FG usage. 

\textit{Contribution of FG}: as described in Section~\ref{Methods}, FG attempts to steer the LLM towards generating correct WPs through program execution feedback. We are interested in observing how and when FG is deployed, and whether or not it can meaningfully guide the LLM in the aforementioned goal.

These criteria will be discussed in the per-configuration performance summaries provided in Section~\ref{overall-results-discussion}.

\subsection{Benchmark Sets}\label{benchmark-sets}
This work utilises four distinct sets of benchmark programs. Each program within these sets is implemented in Java and follows the high-level program structure outlined in Section~\ref{input-programs}. The following points will give a brief overview of each set:

\begin{itemize}
    \item `Existential' benchmarks: this set contains $20$ programs; the correct WPs for the programs of this benchmark set require existential quantification over the array content. For example, a description of a correct WP for some program $P$ in this set might look like -- `\emph{there exists} a valid index $i$ for the input array $a$ such that $a[i]=100$'.
    \item `Universal' benchmarks: this set contains $36$ programs; the correct WPs for the programs of this benchmark set require universal quantification over the array content. For example, a description of a correct WP for some program $P$ in this set could look like -- `\emph{for all} valid indices $i$ of the input array $b$, $b[i]=2i$'.
    \item `Sorting' benchmarks: this set contains $8$ programs. The computations performed in each program feature sorting in some manner. For example, the program in Fig~\ref{fig:program-example} (of Section~\ref{Motivating Example}) features a sorted array in its postcondition.
    \item `Search' benchmarks: this set contains $6$ programs; each of which feature binary-search over an input array. Again, these examples express properties with both existential and universal quantification over the array content. 
\end{itemize}

\subsection{Experimental Set Up and Configurations}\label{set-up}
The set up for our experimental evaluations can be described through the following features:

\begin{itemize}
    \item LLMs: We employ GPT-4o~\cite{gpt-4o-ref} and O4-mini~\cite{o4-mini-ref} from OpenAI. These represent performant and accessible LLMs from the categories of non-reasoning and reasoning models respectively.
    \item Benchmark Sets: The four sets of benchmark programs used are described in Section~\ref{benchmark-sets}. The structure of these programs follows the description outlined in Section~\ref{input-programs}.
    \item Number of Iterations $k$: This defines the number of times WP-generation occurs for each program in each benchmark set. We fix $k=5$ for all of the experiments described.
    \item Fuzzing Parameters: These parameters influence how FG is deployed. We allow a maximum of $10$ attempts of validity-fuzzing (Section~\ref{val-fuzzing}) for each FG-cycle (Section~\ref{fuzzing-cycles}). We also set a maximum fuzzing time of $10s$ for each validity- and weakness-fuzzing (Section~\ref{weak-fuzzing}) invocation.
\end{itemize}

Overall, the four LLM and FG configurations investigated are: GPT-4o without FG (\textit{GPT-4o}), GPT-4o with FG (\textit{GPT-4o-FG}), O4-mini without FG (\textit{O4-mini}), and O4-mini with FG (\textit{O4-mini-FG}). The labels in parenthesis correspond to the configuration labels in both the plots of Fig.(s)~\ref{fig:'Existential plot'},~\ref{fig:'Universal plot'},~\ref{fig:'Sorting plot'},~\ref{fig:'Search plot'} and Table~\ref{tab:overall_results}.

\subsection{Description of the Results Table and Figures}\label{Correctness}
Before delving into the summaries of the obtained results, this section provides a brief descriptive aid to the figures and tables referenced in Section~\ref{overall-results-discussion}.

\subsubsection{Tables} 
The results of Table~\ref{tab:overall_results} depict a set of performance statistics obtained from $k=5$ iterations of evaluations. The majority of the tables headings are self-explanatory, though we believe some clarification regarding `FG Usage' and `FG Success' is warranted. 

The `FG Usage' statistics denote the min, max, and average  number of programs where FG was employed in an attempt to direct the LLM in the task of WP generation. For example, on the `Universal' benchmark set, \textit{GPT-4o-FG} employed FG for a minimum of 6 programs, a maximum of 10 programs, and an average of 7.8 programs per iteration.

The `FG Success' statistics denote the min, max, and average  number of programs where FG was employed \emph{successfully} -- i.e. where FG enabled the generation of a correct WP. For example, on the `Existential' benchmark set, \textit{GPT-4o-FG} employed FG successfully for a minimum of 5 programs, a maximum of 6 programs, and an average of 5.6 programs per iteration.

From this  table alone, we can already conclude that the use of FG has led to a marked improvement in the generation of correct WPs; this is true across all benchmark sets and LLMs used. For GPT-4o (a non-reasoning model), FG provides the greatest improvements in correctness; while in the case of O4-mini (a reasoning model), FG has enabled the achievement of a perfect correctness score across every iteration.

\subsubsection{Figures}
\begin{figure}[t!]
\centering
\includegraphics[width=7cm, height=7cm]{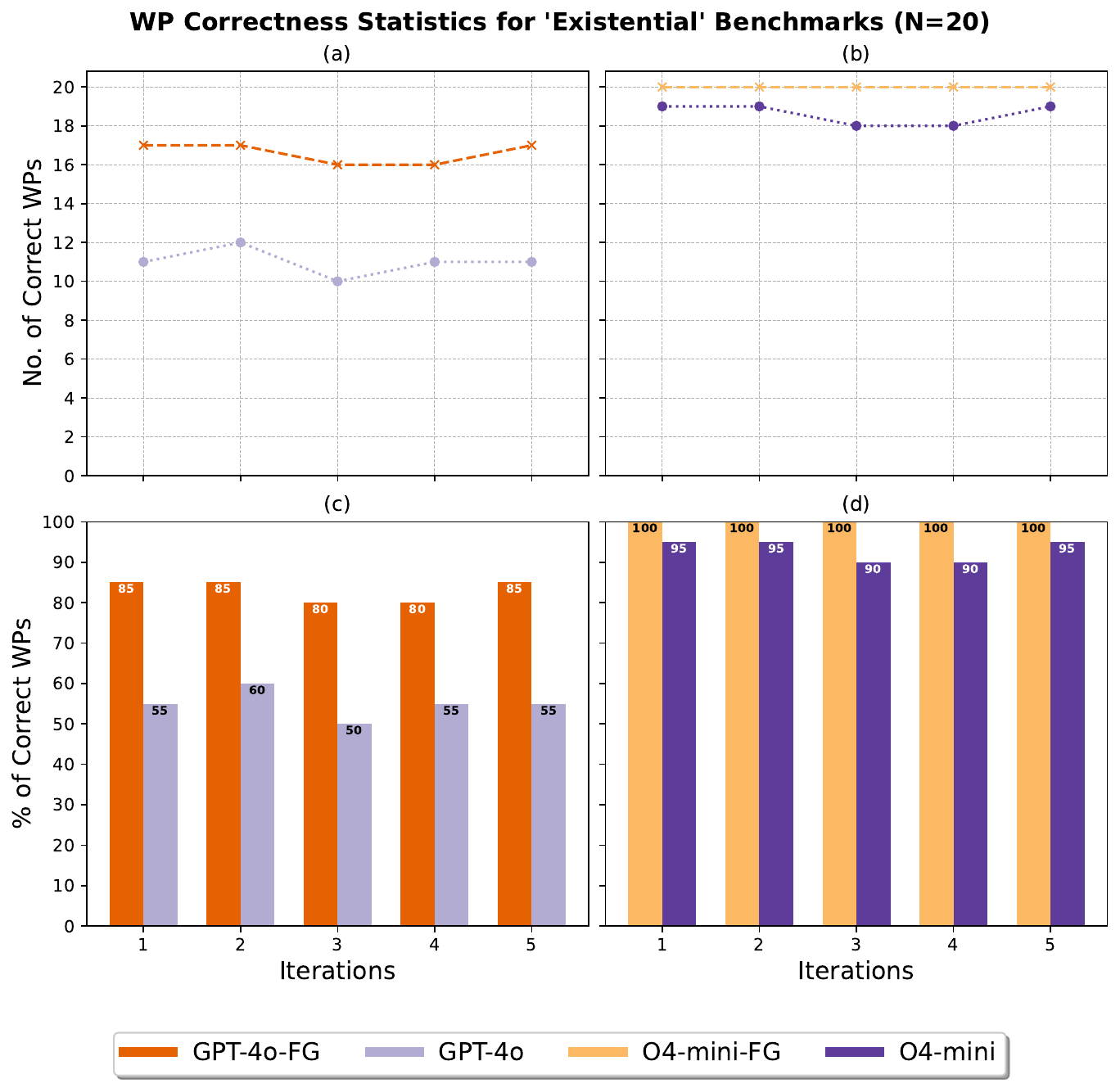}
\caption{WP correctness statistics for the 'Existential' benchmark set, in absolute and relative terms.}
\label{fig:'Existential plot'}
\end{figure}

\begin{figure}[t!]
\centering
\includegraphics[width=7cm, height=7cm]{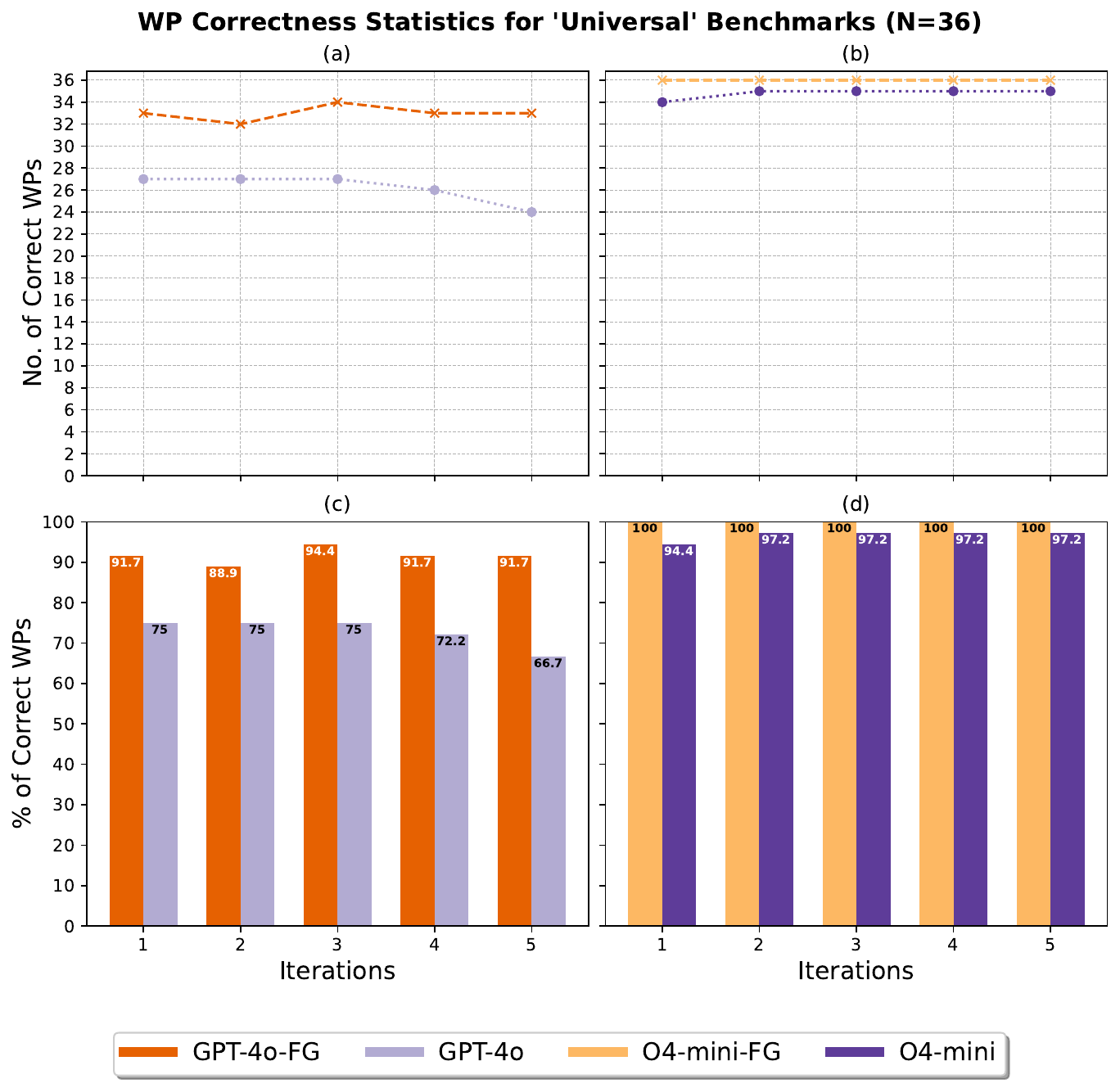}
\caption{WP correctness statistics for the 'Universal' benchmark set, in absolute and relative terms.}
\label{fig:'Universal plot'}
\end{figure}

Fig.(s)~\ref{fig:'Existential plot'},~\ref{fig:'Universal plot'},~\ref{fig:'Sorting plot'},~\ref{fig:'Search plot'} depict the correctness results obtained for each benchmark set and iteration. Each of these figures contains four plots (a)-(d). 

Plots (a) and (b) outline the number of correct WPs generated; these line graphs can aid us in gleaning information about the correctness results achieved per iteration and observe whether or not they are stable. For example, consider the line graph for \textit{GPT-4o-FG} in plot (a) of Fig.~\ref{fig:'Universal plot'}, this shows us that the number of correct WPs generated by this config on the `Universal' benchmark set hovers between 32 and 34 (out of a possible 36) -- indicating result stability.

Plots (c) and (d) portray the same correctness information in terms of percentages; these bar charts enable us to more clearly see the relative performance differences between the experimental configurations. For example, observe the bar chart for \textit{O4-mini} in plot (d) of Fig.~\ref{fig:'Existential plot'}, this shows us that the percentage of correct WPs generated by \textit{O4-mini} on the `Existential' benchmark set always underperforms \textit{O4-mini-FG} in each iteration.

\subsection{Result Summaries per Configuration}\label{overall-results-discussion}

Resulting statistics for each benchmark set and experimental configuration can be seen in Table~\ref{tab:overall_results}. The following provides a summary of these results, along with some observations -- Table~\ref{tab:overall_results} is implicitly referenced throughout.

\subsubsection{GPT-4o without FG (\textit{GPT-4o})}\label{GPT-4o-summary}
This configuration resulted in the lowest average correctness results for all benchmark sets. We can also observe a wide variance here -- with an average correctness percentage of $35\%$ and $72.78\%$ respectively, \textit{GPT-4o} performs poorly at attaining the correct WPs for the `Sorting' benchmarks while performing relatively well on the `Universal' benchmarks. There is also evidence to suggest that the results of this set-up are consistent; this can be seen in the relative similarity of the min and max number of correct WPs generated per benchmark set -- as a result, we can conclude that \textit{GPT-4o} is consistently \textit{incorrect} also. This notion of consistent incorrectness is also evident in plots (a) of Fig.~\ref{fig:'Existential plot'} and Fig.~\ref{fig:'Universal plot'} as the line-graphs appear relatively stable. When we look at the same plots in Fig.~\ref{fig:'Sorting plot'}, and Fig.~\ref{fig:'Search plot'}, we can observe a more erratic performance (exacerbated by the small size of the corresponding benchmark sets).

\subsubsection{GPT-4o with FG (\textit{GPT-4o-FG})}\label{GPT-4o-FG-summary}
Through the use of FG we can see how the performance of the base-line \textit{GPT-4o} config can be significantly improved upon. This is made evident by the fact that \textit{GPT-4o-FG} achieves higher average correctness results than \textit{GPT-4o} across all benchmark sets. Additional evidence that FG is responsible for this performance boost can be seen in the `FG Success' results. For example, the `Sorting' benchmark set sees the largest increase in average correctness (from $35\%$ for \textit{GPT-4o} to $67.50\%$ for \textit{GPT-4o-FG}) and this coincides with the proportionally largest average value for `FG Success'. Another point in support of the efficacy of FG can be seen through the consistent superiority of \textit{GPT-4o-FG} when compared to \textit{GPT-4o}; this is portrayed clearly in plots (a) and (c) of Fig.(s)~\ref{fig:'Existential plot'},~\ref{fig:'Universal plot'},~\ref{fig:'Sorting plot'},~\ref{fig:'Search plot'}

\subsubsection{O4-mini without FG (\textit{O4-mini})} 
\begin{figure}[t!]
\centering
\includegraphics[width=7cm, height=7cm]{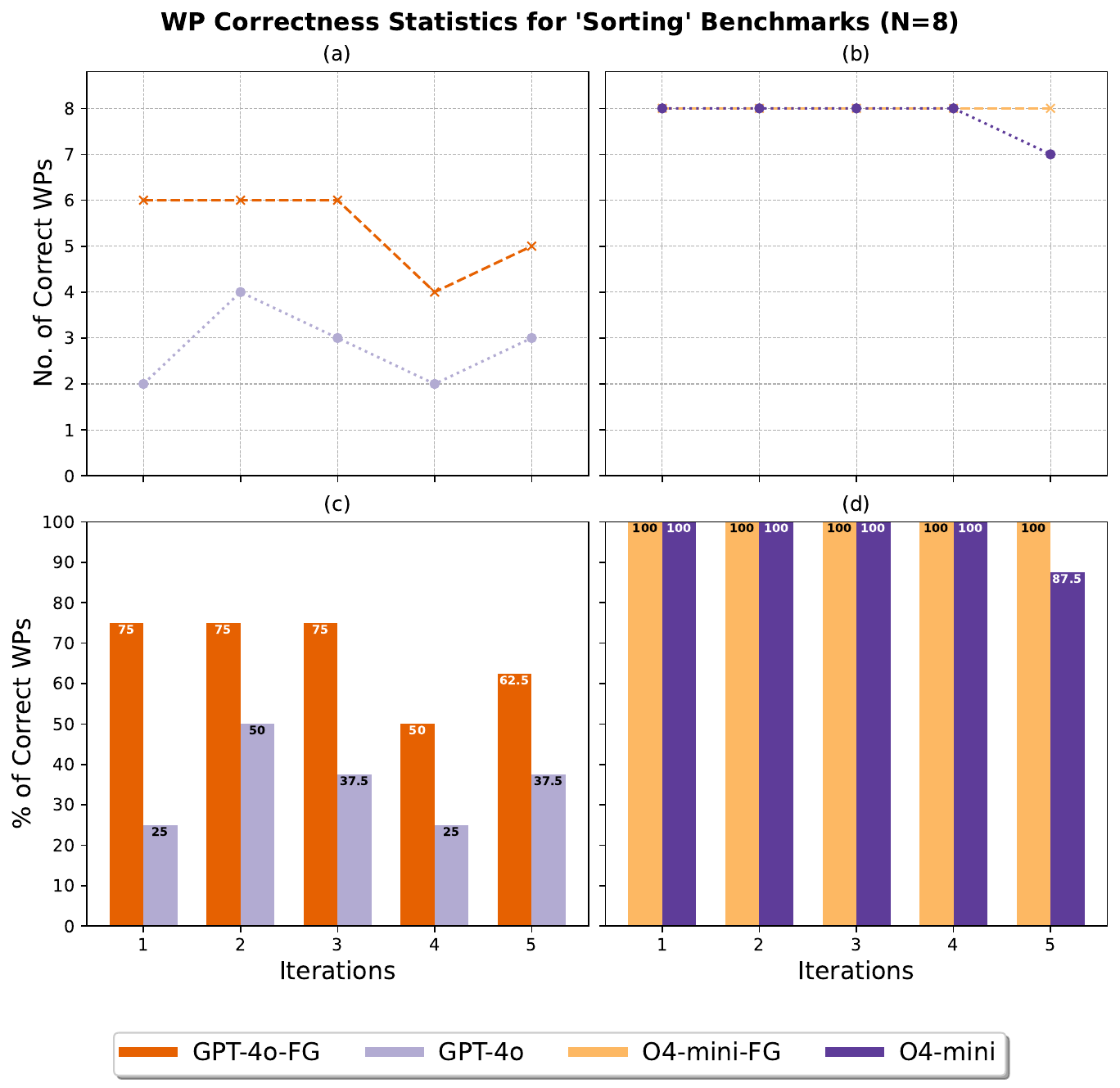}
\caption{WP correctness statistics for the 'Sorting' benchmark set, in absolute and relative terms.}
\label{fig:'Sorting plot'}
\end{figure}

\begin{figure}[t!]
\centering
\includegraphics[width=7cm, height=7cm]{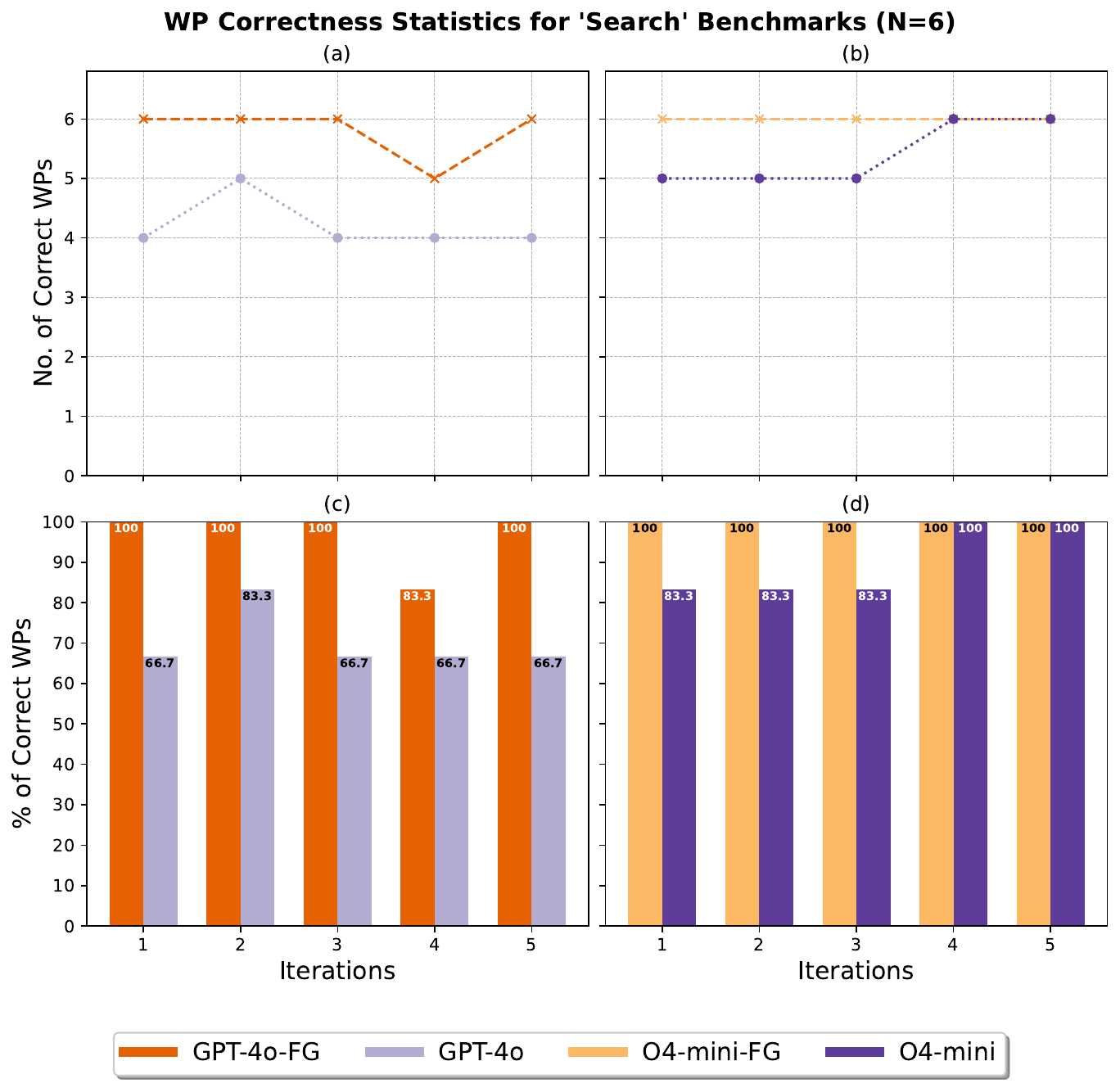}
\caption{WP correctness statistics for the 'Search' benchmark set, in absolute and relative terms.}
\label{fig:'Search plot'}
\end{figure}

Even without the use of FG, \textit{O4-mini} performs exceptionally well in the task of generating correct WPs -- in fact, the lowest average correctness result for \textit{O4-mini} is $90.00\%$ (on the `Search' benchmark set). When considering experimental configurations that do not use FG, the superiority of \textit{O4-mini} in this problem setting is reinforced by observing that the \textit{lowest} average correctness score for \textit{O4-mini} ($90.00\%$ on the `Search' benchmark set) is still higher than the \textit{highest} average correctness score achieved by \textit{GPT-4o} ($72.78\%$ on the `Universal' benchmark set). Although \textit{O4-mini} is dominant over \textit{GPT-4o}, \textit{GPT-4o-FG} can appear to give \textit{O4-mini} a `run for its money' -- this is evidenced by comparable average correctness scores for each benchmark set apart from `Sorting'. This comparable performance between \textit{GPT-4o-FG} and \textit{O4-mini} is further highlighted in each relevant plot of Fig.(s)~\ref{fig:'Existential plot'},~\ref{fig:'Universal plot'},~\ref{fig:'Sorting plot'},~\ref{fig:'Search plot'}.

\subsubsection{O4-mini with FG (\textit{O4-mini-FG})} 
When O4-mini is enabled with FG we can see that the correct WP is attained for every program, in every benchmark set, and for every iteration. Similar to what was observed in \textit{GPT-4o-FG}'s summary (Section~\ref{GPT-4o-summary}), the greatest increase in average correctness (over the corresponding base-line configuration \textit{O4-mini}) also coincides with the largest proportional value of `FG Success'. 

\subsection{Notable Cases}\label{notable-cases}

\subsubsection{GPT-4o Can't Swap}
This case concerns three programs from the `Existential' benchmark set and the \textit{GPT-4o} configuration. These three programs each perform a value swap between two variables in their respective loop computations (see the supplementary material link for example-two in Section~\ref{supp-mat}). In each instance, \textit{GPT-4o} produced a candidate WP which failed to account for this swapping of values -- this was true for every evaluation carried out; in the end, \textit{GPT-4o} never generated a correct WP for these programs.

Another interesting point regarding the aforementioned programs is that FG generally did \emph{not} help. The WPs that were required by these programs postconditions were complex enough that validity-fuzzing (Section~\ref{val-fuzzing}) could not produce any corresponding inputs for testing. This problem arises due to the random-input generators of the underlying fuzz testing tool JQF~\cite{padhye2019jqf}. This phenomena resulted in validity-fuzzing phases `passing' with invalid candidate WPs. Here, the combined inadequacies of GPT-4o and FG led to a perfect storm for poor performance.

\subsubsection{O4-mini's Trouble with Integer Arithmetic}
This example regards a program in the `Universal' benchmark set (see the supplementary material link for example-three in Section~\ref{supp-mat}). Here, O4-mini produced a candidate WP that is correct for mathematical arithmetic, but that is actually invalid when considering integer arithmetic in Java. The initial WP generated was susceptible to integer overflow and this flaw was picked up by the validity-fuzzing phase of FG. This feedback was then given to O4-mini and the LLM could then produce the correct WP; this behaviour was observed in each iteration carried out.

\section{Discussion}\label{Discussion}
\subsection{Threats to Validity}
The main threats to validity arise from our use of LLMs. The outputs of LLMs are inherently stochastic in nature~\cite{vaswani2017attention}, and so the results that we have derived may not be \emph{exactly} reproducible. We attempt to mitigate the severity of this threat by providing the prompts and logs of our interactions with the OpenAI API.

Another threat arises when we consider the range of LLMs employed. This work solely features models from a single vendor (OpenAI) and therefore the results that we have produced are biased in this respect; the argument could be made that other models may perform significantly better or worse, and that our results do not represent this potential variety.

Additionally, data leakage~\cite{inan2021training} is typically a concern for many studies which rely on LLMs. Data leakage occurs when the information from a test set (e.g. benchmark programs) is already present in the training data of an LLM. We believe we have significantly reduced the risk of this threat by creating our benchmark programs from scratch, and hosting them in private repositories.

%\subsection{Limitations}
As is common with many empirical studies in software engineering, our work is inherently limited by the scope of the experimental benchmarks used. We attempted to alleviate this concern by ensuring that our benchmark programs covered a range of common programming operations and tasks, e.g. arithmetic, sorting, search, looping computations, etc.
However, we have only explored relatively small programs operating over integer arrays. A more extensive experimentation with real code is necessary to support the practical benefits of our technique, which we leave to future work.

\section{Related Work}\label{Related Work}
The combination of LLMs with formal methods and additional software analysis techniques is leading to a burgeoning sub-field of computer science.

Recent works have employed LLMs with formal tools to generate program properties that are adjacent to weakest preconditions (WPs); much of this work has revolved around invariant generation -- a key component for software verification. The authors of~\cite{pirzada2024llm} use LLMs to generate loop invariants that are then validated by the Vampire theorem-prover~\cite{kovacs2013first}; these validated invariants are then used to replace loops within the context of bounded model checking~\cite{clarke2001bounded}. The work of~\cite{wu2023lemur} introduces the Lemur framework where LLMs are used to generate and repair candidate invariants that are then used within the ESBMC~\cite{menezes2024esbmc} and UAtomizer~\cite{heizmann2016ultimate} frameworks. The authors of~\cite{pei2023can} fine-tune LLMs in an attempt to infer invariants for Java programs (such that these invariants are likely to be Daikon~\cite{ernst2007daikon} generated). In a task adjacent to invariant generation, the authors of~\cite{endres2024can} use LLMs to generate postconditions from natural language documentation. Additionally,~\cite{si2020code2inv} does not use LLMs, but instead opts to use deep-reinforcement-learning to synthesize program invariants. 

The task of generating preconditions has also been approached on several fronts~\cite{abstract-debugging93,footprint-analysis07,CousotEtal02,procedure-summaries07,modular-assertion-checking08,astorga2019learning,cousot2011precondition,gehr2015learning,Moy08,padhi2016data,sankaranarayanan2008dynamic,SeghirEtal13,sumanth2024maximal,ZhouEtal21,sumanth2024weakest}. The authors of~\cite{sumanth2024weakest} and~\cite{sumanth2024maximal} use logical abductive inference to infer the provably weakest preconditions for a class of deterministic and non-deterministic array programs. The works of~\cite{padhi2016data} and~\cite{sankaranarayanan2008dynamic} employ a more dynamic approach, utilising data-driven execution loops in order to learn likely preconditions. Our work differs from the above in that we combine fuzzing and LLMs in a unique manner to derive WPs; additionally, our approach can handle problem variants (including sorting and search), that the aforementioned techniques cannot (see the supplementary material link for example-four in Section~\ref{supp-mat}).

\section{Future Work}\label{Future Work}
There are several avenues for future work which arise from this current study, a number of which stem from the limitations of our current approach towards generating weakest preconditions (WPs). For example, the WPs that are derived through LLMs and Fuzzing Guidance (FG) are only ever \emph{likely}-valid and \emph{likely}-weakest (see Section~\ref{fuzz-guidance-phases}); this is due to the inherently incomplete nature of fuzz testing (and thus FG). One direction for future work would seek to concretely verify the weakness and validity of a candidate WP that had been generated through an LLM with FG; this could be done through the use of a formal reasoning tool like Vampire~\cite{kovacs2013first}.

Another path for future work would seek to expand the capabilities of the fuzz testing tool which undergirds FG -- namely, JQF~\cite{padhye2019jqf}. This work might include creating a more adept and flexible orchestrator for FG, such that FG could be applied to a greater range of programs beyond those studied here.

Bug-finding is another application of the approach outlined in this work. Anecdotally, we found that WPs could be generated for real-world programs which were similar to the benchmarks here; these WPs were then able to avoid known exception-triggering test cases and reveal other fault-inducing test cases that were not already known. We also found that buggy programs could be revealed in a roundabout manner when the WP generated by an LLM with FG does not fit with a programmer's expected WP.

\section{Conclusion}\label{Conclusions}
This work investigated the use of LLMs (specifically GPT-4o and O4-mini) for generating weakest preconditions (WPs) and introduced our novel mechanism of Fuzzing Guidance (FG) for aiding LLMs towards this goal. The effectiveness of our approach was demonstrated on a comprehensive benchmark set of deterministic array programs in Java. 

In concrete terms, we found that O4-mini was generally superior to GPT-4o in generating correct WPs. Additionally, we observed that FG enabled O4-mini to consistently generate the correct WPs for all of the benchmark programs studied. Our results also showed that a less performant model such as GPT-4o, when enabled with FG, could achieve competitive performance relative to the baseline O4-mini model (i.e. O4-mini without FG).

Overall, our results indicate that LLMs are capable of producing viable candidate WPs, and that this ability is enhanced through FG.

\backmatter

\bmhead{Supplementary material}\label{supp-mat} Log files detailing GPT interactions and fuzzing invocations: https://doi.org/10.5281/zenodo.17018494

\bmhead{Acknowledgements}

This work was conducted with the financial support of the Research Ireland Centre for Research Training in Digitally-Enhanced Reality (d-real) under Grant No. 18/CRT/6224, Taighde Éireann – Research Ireland under Grant number 13/RC/2094/\textbackslash\_2, and the ERC Consolidator Grant ARTIST 101002685. 
For the purpose of Open Access, the author has applied a CC BY public copyright licence to any Author Accepted Manuscript version arising from this submission.


\bibliography{sn-bibliography}% common bib file

\end{document}